\documentstyle[amssymb,twocolumn,aps]{revtex}

\begin{document}
\title{Microscopic mechanisms of magnetization reversal}
\author{Vladimir L. Safonov}
\address{Center for Magnetic Recording Research, \\
University of California -San Diego, \\
9500 Gilman Drive, La Jolla, CA 92093-0401}
\date{\today }
\maketitle

\begin{abstract}
Two principal scenarios of magnetization reversal are considered. In the
first scenario all spins perform coherent motion and an excess of magnetic
energy directly goes to a nonmagnetic thermal bath. A general dynamic
equation is derived which includes a tensor damping term similar to the
Bloch-Bloembergen form but the magnetization magnitude remains constant for
any deviation from equilibrium. In the second reversal scenario, the
absolute value of the averaged sample magnetization is decreased by a rapid
excitation of nonlinear spin-wave resonances by uniform magnetization
precession. We have developed an analytic ${\bf k}$-space micromagnetic
approach that describes this entire reversal process in an ultra-thin soft
ferromagnetic film for up to $90^{o}$ deviation from equilibrium. Conditions
for the occurrence of the two scenarios are discussed.
\end{abstract}


\section{Introduction}

Studies of magnetization reversal in ultra-thin ferromagnetic films under an
applied external magnetic field are of great importance in magnetic
recording physics. A conventional theoretical tool to study magnetization
reversal is based on the phenomenological Landau-Lifshitz equation \cite{LL}
or, its equivalent modification with the Gilbert form of relaxation \cite%
{gilbert}. These equations conserve the absolute value of magnetization ($%
\left\vert {\bf M}\right\vert ={\rm const}$){\rm \ }in a single domain
region. Both equations were introduced (a) for small magnetization motions
and (b) for the case of uniaxial magnetic symmetry. The energy losses are
defined by an isotropic phenomenological damping fitting parameter $\alpha $
(\textquotedblleft damping constant\textquotedblright ).

Recently a theoretical approach \cite{safonov}, \cite{safbertbook}, \cite%
{wangbertsaf1}, \cite{BertSafJin}, \cite{nonrelax} has been developed to
correct the limitations of the Landau-Lifshitz-Gilbert (LLG) theory. The
main idea was to represent the magnetization dynamics as the motion of a
damped nonlinear oscillator with the random force of thermal fluctuations.
The oscillator model is a convenient tool to establish a \textquotedblleft
bridge\textquotedblright\ between the microscopic physics, where the
rotational oscillator complex variables naturally describe spin excitations
and the macroscopic magnetization dynamics. It has been rigorously shown by
including specific coupling of a magnetic system to a variety of loss
mechanisms \cite{safbertrelax}, \cite{BertWang} that for small oscillations
near equilibrium the macroscopic damping term reflects the anisotropy of the
system.

Our aim in this paper is to develop a self-consistent picture that describes
the entire reversal process. We consider two possible scenarios: 1) the
magnetization reverses uniformly, involving nonlinear dynamic damping; 2)
the magnetization reverses nonuniformly, involving the excitation of
nonlinear spin waves. We give an explicit criterion for this nonuniform
process for an untra-thin magnetic film.

{\it Scenario \#1:}{\em \ }the total film magnetization $|{\bf M}|$ is
constant during reversal. All spins perform a coherent motion (the role of
non-uniform spin motions is neglected). An excess of magnetic energy goes
directly to a nonmagnetic thermal bath. We derive a general magnetization
dynamic equation from a nonlinear oscillator model. The nonlinear damping
follows from the variety of well-known physical damping mechanisms \cite%
{nonrelax}, \cite{safbertrelax}. Here we extend our previous results for
uniaxial symmetry to the case of non-uniaxial symmetry.

{\it Scenario \#2:} the total film magnetization $|{\bf M}|$ decreases.
Experimentally it was observed in Ref. \cite{SilvaKabosPufall}. Large angle
magnetization motion can excite spin-wave instabilities, which increase
substantially the magnetization reversal rate \cite{suhl2}. We explicitly
evaluate the second order Suhl instability and construct a self-consistent
theory of magnetization switching for up to $90^{o}$ from equilibrium.

The problem of nonlinear spin-wave excitation during reversal has been
explored by numerical simulations in nanograins \cite{safbert}, \cite%
{safbert1}, and thin films \cite{zhufilm}, \cite{boerner}. All these
simulations have been performed using conventional local micromagnetic
modeling, which includes: a) the analysis of intra- and inter-cell
interactions, b) analysis of phenomenological dynamic equations, and c)
computer simulations. There are two principal problems in this technique: 1)
the physical problem of the introduction of local phenomenological damping
(and corresponding magnetic noise) and 2) the computing problem in the case
of a large number of cells.

Both problems of local micromagnetic modeling can be avoided by developing
the ${\bf k}$-space micromagnetic modeling as we do in this work. Our theory
includes: a) an analysis of spin-wave spectra and interactions in an
ultra-thin film, b) the calculation of the effective scattering processes
(most of the accumulated energy is to be transformed to nonlinear spin
waves), c) the analysis of self-consistent dynamic equations with
microscopic damping (and noise, if necessary). Note that a similar technique
to study nonlinear spin-wave dynamics has been developed in the theory of
parametric magnon excitation (mainly in the bulk, see, e.g., \cite%
{schloemann}, \cite{zakharov}, \cite{lvov}, \cite{Safpair}). We have already
considered ${\bf k}$-space modeling in application to magnetic noise in a
thin film \cite{safbertnonuninoise}. Recently Dobin and Victora \cite%
{dobinvictora} estimated the increment of the second order Suhl instability
and corresponding effective magnetization reversal time for up to $25^{o}$
deviation from equilibrium in the film plane. Here we give an explicit
analytic formulation to describe magnetization reversal (switching) for up
to $90^{o}$ deviation from equilibrium in ultra-thin films in terms of
spin-wave pair excitations.

\section{Scenario \#1: $\left\vert {\bf M}\right\vert ={\rm const}$}

In this section we consider the magnetization reversal without spin-wave
excitations. The approach is to transform the magnetization dynamics without
damping to normal mode coordinates. Then we introduce nonlinear damping,
which has a connection with microscopic physics and return back to
magnetization coordinates. The analysis parallels the approach for low-level
excitations.

Let us consider small-amplitude magnetization motions of a single-domain
ferromagnetic particle in the vicinity of equilibrium state ${\bf M}||%
\widehat{{\bf z}}_{0}$, where $\widehat{{\bf z}}_{0}$ is the unit vector in
the equilibrium direction. The magnetization rotation around effective field
in this case, in general, is elliptical and the magnetic energy ${\cal E}$
can be represented as a quadratic form:

\begin{equation}
{\cal E}/\left( \frac{M_{s}V}{\gamma }\right) =\frac{\gamma H_{x_{0}}}{2}%
m_{x_{0}}^{2}+\frac{\gamma H_{y_{0}}}{2}m_{y_{0}}^{2}.  \label{lin energy}
\end{equation}
Here ${\bf m}\equiv {\bf M}/M_{s}$, $\widehat{{\bf x}}_{0}$ and $\widehat{%
{\bf y}}_{0}$ are the unit orthogonal vectors in the plane perpendicular to
the equilibrium direction, $M_{s}$ is the saturation magnetization and $V$
is the particle volume. $H_{x_{0}}$ and $H_{y_{0}}$\ are positive Kittel
\textquotedblleft stiffness\textquotedblright\ fields, which include both
microscopic and shape anisotropies and the external magnetic field. The
parameter $M_{s}V/\gamma \equiv \hbar S$, where $\hbar $ is Planck's
constant and $S$ is the total spin of the film.

From the Holstein-Primakoff transformation \cite{hopri} we have:

\begin{eqnarray}
m^{+} &=&a\sqrt{1+m_{z_{0}}},\quad m_{z_{0}}=1-a^{\ast }a,  \label{hoprim} \\
m^{-} &=&a^{\ast }\sqrt{1+m_{z_{0}}},\quad m^{\pm }=m_{x_{0}}\pm im_{y_{0}},
\nonumber
\end{eqnarray}
where $a^{\ast }$ and $a$ describe spin excitations.

The magnetic energy ({\ref{lin energy}}) can be written in the quadratic
form:

\begin{equation}
{\cal E}/\left( \frac{M_{s}V}{\gamma }\right) ={\cal A}a^{\ast }a+\frac{%
{\cal B}}{2}(aa+a^{\ast }a^{\ast }),  \label{quadform}
\end{equation}
where ${\cal A}=\gamma (H_{x_{0}}+H_{y_{0}})/2$ and ${\cal B}=\gamma
(H_{x_{0}}-H_{y_{0}})/2$. The non-diagonal terms in (\ref{quadform}) are
eliminated by the linear canonical transformation (e.g., \cite{sparks}): 
\begin{eqnarray}
a=uc+vc^{\ast }, &\quad &a^{\ast }=uc^{\ast }+vc,  \label{canonic} \\
u=\sqrt{\frac{{\cal A}+\omega _{0}}{2\omega _{0}}}, &\quad &v=-\frac{{\cal B}%
}{|{\cal B}|}\sqrt{\frac{{\cal A}-\omega _{0}}{2\omega _{0}}}.  \nonumber
\end{eqnarray}
The energy in terms of the normal mode coordinates $c$ and $c^{\ast }$ is
simply: 
\begin{equation}
{\cal E}/\left( \frac{M_{s}V}{\gamma }\right) =\omega _{0}c^{\ast }c,
\label{ener}
\end{equation}
where $\omega _{0}=\sqrt{{\cal A}^{2}-{\cal B}^{2}}=\gamma \sqrt{%
H_{x_{0}}H_{y_{0}}}$ is the ferromagnetic resonance frequency.

The dynamic equations for $c$ and $c^{\ast }$ are independent and can be
written as:

\begin{equation}
\frac{dc}{dt}+\eta c=-i\omega _{0}c,\quad \frac{dc^{\ast }}{dt}+\eta c^{\ast
}=i\omega _{0}c^{\ast }.  \label{c_equations}
\end{equation}
Here $\eta $ is the linear relaxation rate, which can be found
microscopically \cite{safbertrelax}. In the case of large magnetization
motion we can write a corresponding nonlinear oscillator equation in the
form:

\begin{equation}
\frac{dc}{dt}+\eta (N)c=G(c,c^{\ast }),\quad N\equiv c^{\ast }c.
\label{c nonlinear}
\end{equation}%
Here $G(c,c^{\ast })$ corresponds to the gyromagnetic term $-\gamma {\bf m}%
\times {\bf H}_{{\rm eff}}$. The nonlinear relaxation rate $\eta (N)$ can be
estimated from the known relaxation process for the uniform precession \cite%
{nonrelax}. We assume that $\left\vert {\bf m}\right\vert =1$ and therefore
no spin waves are excited. 

Using back transformations (\ref{canonic}) and (\ref{hoprim}), we can derive
an equation (corresponding to (\ref{c nonlinear})) in terms of ${\bf m}$%
-components:

\begin{eqnarray}
\frac{d{\bf m}}{dt} &=&-\gamma {\bf m}\times {\bf H}_{{\rm eff}}-\stackrel{%
\leftrightarrow }{\Gamma }\cdot ({\bf m-}\widehat{{\bf z}}_{0}),
\label{basic equation 1} \\
\stackrel{\leftrightarrow }{\Gamma } &=&2\eta (N)\left( 
\begin{array}{ccc}
\frac{m_{z_{0}}}{1+m_{z_{0}}} & 0 & 0 \\ 
0 & \frac{m_{z_{0}}}{1+m_{z_{0}}} & 0 \\ 
0 & 0 & 1%
\end{array}%
\right) ,  \nonumber
\end{eqnarray}%
where 
\begin{equation}
N=\frac{{\cal A}}{\omega _{0}}(1-m_{z_{0}})+\frac{{\cal B}}{\omega _{0}}%
\frac{m_{x_{0}}^{2}-m_{y_{0}}^{2}}{1+m_{z_{0}}}.  \label{n}
\end{equation}%
Note that the Eq.(\ref{basic equation 1}) conserves the magnitude of ${\bf m}
$. For small deviations from equilibrium, when $m_{z_{0}}\simeq 1$, this
equations exactly correspond to Bloch-Bloembergen equations \cite{BB} with $%
\eta (0)=1/T_{2}$ and $1/T_{1}=2/T_{2}$.\ From Eq.(\ref{basic equation 1})
we see that one can expect the anomalously large damping in the case of $%
180^{o}$ reversal when $1+m_{z_{0}}\rightarrow 0$ (see, also \cite{nonrelax}%
, \cite{wangbert}). Simple numerical analysis shows that magnetization
dynamics in the framework of Eq.(\ref{basic equation 1}) for angles about $%
70^{o}$ (and smaller) can be approximated by LLG dynamics. This can explain
why the switching experiment \cite{SilvaKabosPufall} with $\left\vert {\bf M}%
\right\vert ={\rm const}$ was well fitted by LLG equation.

\section{Scenario \#2: $\left\vert {\bf M}\right\vert \neq {\rm const}$}

We shall consider an ultra-thin ferromagnetic film ($\tau \times L_{y}\times
L_{z}$) with the magnetization: ${\bf M}({\bf r})=M_{s}{\bf m}({\bf r})$, $%
{\bf r}=(y,z)$. The variation of the unit vector ${\bf m}$ within the film
thickness ($-\tau /2\leq x\leq \tau /2$) will be neglected. Locally one has: 
$\left| {\bf m(r)}\right| {\bf =}\sqrt{m_{x}^{2}+m_{y}^{2}+m_{z}^{2}}=1$.

In order to introduce collective magnetization motions, we assume that the
film is periodic along both $y$ and $z$ directions with periods $L_{y}$ and $%
L_{z}$, respectively. The Fourier series representation can be written as

\begin{eqnarray}
{\bf m}({\bf r}) &=&%
\mathop{\displaystyle\sum}%
\limits_{{\bf k}}{\bf m}_{{\bf k}}\exp (i{\bf k\cdot r}),  \label{Fourier} \\
{\bf m}_{{\bf k}} &=&\frac{1}{L_{y}L_{z}}%
\displaystyle\int %
\limits_{0}^{L_{y}}dy%
\displaystyle\int %
\limits_{0}^{L_{z}}dz\ {\bf m}({\bf r})\exp (-i{\bf k\cdot r}),  \nonumber
\end{eqnarray}
$k_{y}=2\pi n_{y}/L_{y}$, $k_{z}=2\pi n_{z}/L_{z}$ ($-\infty
<n_{y},n_{z}<\infty $) are the wave vector components in the plane.

The equilibrium is supposed to be a uniformly magnetized state, in which the
magnetization is oriented in the $(y,z)$ plane along an equilibrium axis $%
z_{0}$. The transformation from the $(x,y,z)$ coordinates to equilibrium
coordinates (Fig.1) $(x_{0},y_{0},z_{0})$ is defined by

\begin{equation}
\left( 
\begin{array}{c}
y \\ 
z%
\end{array}
\right) =\left( 
\begin{array}{cc}
\cos \theta _{0} & \sin \theta _{0} \\ 
-\sin \theta _{0} & \cos \theta _{0}%
\end{array}
\right) \left( 
\begin{array}{c}
y_{0} \\ 
z_{0}%
\end{array}
\right)  \label{transform}
\end{equation}
Here $\theta _{0}\,$determines a rotation in the film plane, $x=x_{0}$.
Analogous transformation should be used for $(m_{y},m_{z})\rightarrow
(m_{y_{0}},m_{z_{0}})$\ and wave vector components $(k_{y},k_{z})\rightarrow
(k_{y_{0}},k_{z_{0}})$. Note that both the absolute value of the wave vector 
$k=\left\vert {\bf k}\right\vert $ and the scalar product ${\bf k\cdot r}$
are invariant in respect to choice of system of coordinates.

\subsection{Magnetic energy}

The magnetic energy of the film contains the exchange energy, energy of
anisotropy, Zeeman energy and demagnetization energy: 
\begin{equation}
{\cal E}={\cal E}_{exch}+{\cal E}_{anis}+{\cal E}_{Z}+{\cal E}_{dmag}.
\label{energy}
\end{equation}

The exchange energy $-A({\bf \bigtriangledown \cdot m})^{2}$ can be
represented as

\begin{equation}
{\cal E}_{exch}=VA%
\mathop{\displaystyle\sum}%
\limits_{{\bf k}}k^{2}\left( m_{y_{0},{\bf k}}m_{y_{0},-{\bf k}}+m_{z_{0},%
{\bf k}}m_{z_{0},-{\bf k}}\right) ,  \label{k-exchange}
\end{equation}
where $A$ is the exchange constant and $V=\tau L_{y}L_{z}$ is the film
volume. To obtain (\ref{k-exchange}) we have used the following formula:

\begin{equation}
\frac{1}{L_{y}L_{z}}%
\displaystyle\int %
\limits_{0}^{L_{y}}dy%
\displaystyle\int %
\limits_{0}^{L_{z}}dz\ \exp [i({\bf k}+{\bf k}_{1}){\bf \cdot r}]=\Delta (%
{\bf k}+{\bf k}_{1}),  \label{formula delta}
\end{equation}
where $\Delta (\cdot )$ is the Kronecker delta function: $\Delta ({\bf q})=1$%
, if ${\bf q}=0$ and $\Delta ({\bf q})=0$ otherwise.

The quadratic uniaxial anisotropy energy ($z$ is an easy axis, see, Fig. 1)
in the ${\bf k}$-space is:

\begin{eqnarray}
{\cal E}_{anis} &=&-VK_{1}%
\mathop{\displaystyle\sum}%
\limits_{{\bf k}}\left( -m_{y_{0},{\bf k}}\sin \theta _{0}+m_{z_{0},{\bf k}%
}\cos \theta _{0}\right)  \nonumber \\
&&\times \left( -m_{y_{0},-{\bf k}}\sin \theta _{0}+m_{z_{0},-{\bf k}}\cos
\theta _{0}\right) .  \label{energy-anisotropy}
\end{eqnarray}

The Zeeman energy in the external magnetic field ${\bf H}_{0}=(0,\ H_{0}\sin
\theta _{H},\ H_{0}\cos \theta _{H})$ is:

\begin{equation}
{\cal E}_{Z}=-VM_{s}H_{0}[m_{y_{0},0}\sin (\theta _{H}-\theta
_{0})+m_{z_{0},0}\cos (\theta _{H}-\theta _{0})].  \label{zeem k}
\end{equation}

The demagnetization energy (see, \cite{mansuripur}) is defined by :

\begin{eqnarray}
&&{\cal E}_{dmag}=2\pi M_{s}^{2}V%
\mathop{\displaystyle\sum}%
\limits_{{\bf k}}\{G(k\tau )m_{x_{0},{\bf k}}m_{x_{0},-{\bf k}}
\label{demag equi} \\
&&+[1-G(k\tau )][\left( \frac{k_{y_{0}}}{k}\right) ^{2}m_{y_{0},{\bf k}%
}m_{y_{0},-{\bf k}}  \nonumber \\
&&+\left( \frac{k_{z_{0}}}{k}\right) ^{2}m_{z_{0},{\bf k}}m_{z_{0},-{\bf k}%
}+2\frac{k_{y_{0}}k_{z_{0}}}{k^{2}}m_{y_{0},{\bf k}}m_{z_{0},-{\bf k}}]\}, 
\nonumber
\end{eqnarray}
where $G(x)=[1-\exp (-x)]/x$.

\subsection{Spin waves}

We shall utilize a classical form of the spin representation in terms of
Bose operators introduced in Refs. \cite{villain}, \cite{baryab} and
convenient for 2D systems. For the unit magnetization vector ${\bf m}$ this
representation in terms of complex variables $a$ and $a^{\ast }$ can be
written as:

\begin{mathletters}
\begin{eqnarray}
m_{x_{0}} &=&i\frac{a-a^{\ast }}{\sqrt{2}},  \label{mx} \\
m_{y_{0}} &=&\sqrt{1-m_{x_{0}}^{2}}\sin \left( \frac{a+a^{\ast }}{\sqrt{2}}%
\right) ,  \label{my} \\
m_{z_{0}} &=&\sqrt{1-m_{x_{0}}^{2}}\cos \left( \frac{a+a^{\ast }}{\sqrt{2}}%
\right) .  \label{mz}
\end{eqnarray}
An expansion of (\ref{mx})-(\ref{mz}) up to the fourth order gives accuracy $%
\sim 6\%$ for about $90^{o}$ deviation from equilibrium:

\end{mathletters}
\begin{mathletters}
\begin{eqnarray}
m_{x_{0}} &=&i\frac{a-a^{\ast }}{\sqrt{2}},  \label{mx1} \\
m_{y_{0}} &\simeq &\frac{a+a^{\ast }}{\sqrt{2}}+\frac{a^{3}+(a^{\ast
})^{3}-3a^{\ast }a^{2}-3(a^{\ast })^{2}a}{6\sqrt{2}},  \label{my1} \\
m_{z_{0}} &\simeq &1-a^{\ast }a-\frac{a^{4}+(a^{\ast })^{4}-2a^{\ast
}a^{3}-2(a^{\ast })^{3}a}{12}.  \label{mz1}
\end{eqnarray}

The following Fourier representation for $a({\bf r})$ (and its complex
conjugate) will be used:

\end{mathletters}
\begin{eqnarray}
a({\bf r}) &=&\sum\limits_{k}a{_{{\bf k}}}\exp (i{\bf kr}_{j}),  \label{ak}
\\
a{_{{\bf k}}} &=&\frac{1}{L_{y}L_{z}}%
\displaystyle\int %
\limits_{0}^{L_{y}}dy%
\displaystyle\int %
\limits_{0}^{L_{z}}dz\ a({\bf r})\exp (-i{\bf k\cdot r}).  \nonumber
\end{eqnarray}

In general, the magnetic energy can be expressed as

\begin{equation}
{\cal E}={\cal E}^{(0)}+{\cal E}^{(1)}+{\cal E}^{(2)}+{\cal E}^{(3)}+{\cal E}%
^{(4)}+...,  \label{energy 01234}
\end{equation}
where the superscript denotes an order in terms of $a$ and $a^{\ast }$.

The zeroth order energy term is equal to

\begin{equation}
{\cal E}^{(0)}=-VK_{1}\cos ^{2}\theta _{0}-VM_{s}H_{0}\cos (\theta
_{H}-\theta _{0}).  \label{zeroth order}
\end{equation}
The equilibrium uniformly magnetized state is defined by the condition: $%
\partial {\cal E}^{(0)}/\partial \theta _{0}=0$, which corresponds to

\begin{equation}
H_{K}\sin 2\theta _{0}=2H_{0}\sin (\theta _{H}-\theta _{0}).  \label{equi}
\end{equation}
Here $\,H_{K}=2K_{1}/M_{s}\,$ is the anisotropy field. In order to have a
stable stationary state, we need $\partial ^{2}{\cal E}^{(0)}/\partial
\theta _{0}^{2}>0$. The first order energy term ${\cal E}^{(1)}=0$ at the
equilibrium and this condition coincides with Eq.(\ref{equi}).

The quadratic term has the form:

\begin{equation}
{\cal E}^{(2)}/\left( \frac{M_{s}V}{\gamma }\right) =\sum\limits_{{\bf k}}%
\left[ {\cal A}_{{\bf k}}a_{{\bf k}}^{\ast }a_{{\bf k}}+\frac{{\cal B}{_{%
{\bf k}}}}{2}(a_{{\bf k}}a_{-{\bf k}}+a_{{\bf k}}^{\ast }a_{-{\bf k}}^{\ast
})\right] ,  \label{quadratic form}
\end{equation}
where

\begin{eqnarray}
{\cal A}_{{\bf k}} &=&\gamma \alpha _{E}k^{2}-\frac{\gamma H_{K}}{2}\sin
^{2}\theta _{0}+\gamma H_{0}\cos (\theta _{H}-\theta _{0})  \nonumber \\
&&+2\pi \gamma M_{s}\left[ [1-G(k\tau )]\left( \frac{k_{y_{0}}}{k}\right)
^{2}+G(k\tau )\right] ,  \label{Ak}
\end{eqnarray}
\begin{eqnarray}
{\cal B}{_{{\bf k}}} &=&\gamma \alpha _{E}k^{2}-\frac{\gamma H_{K}}{2}\sin
^{2}\theta _{0}  \nonumber \\
&&+2\pi \gamma M_{s}\left[ [1-G(k\tau )]\left( \frac{k_{y_{0}}}{k}\right)
^{2}-G(k\tau )\right] ,  \label{Bk}
\end{eqnarray}
and $\alpha _{E}\equiv A/M_{s}$. Using the following linear canonical
transformation (e.g., \cite{sparks}):

\begin{equation}
a_{{\bf k}}=u_{{\bf k}}c_{{\bf k}}+v_{{\bf k}}c_{-{\bf k}}^{\ast },\quad a_{%
{\bf k}}^{\ast }=u_{{\bf k}}c_{{\bf k}}^{\ast }+v_{{\bf k}}c_{-{\bf k}},
\label{u-v transform}
\end{equation}

\begin{equation}
u_{{\bf k}}=\sqrt{\frac{{\cal A}_{{\bf k}}+\omega _{{\bf k}}}{2\omega _{{\bf %
k}}}},\quad v_{{\bf k}}=-\frac{{\cal B}_{{\bf k}}}{|{\cal B}_{{\bf k}}|}%
\sqrt{\frac{{\cal A}_{{\bf k}}-\omega _{{\bf k}}}{2\omega _{{\bf k}}}},
\label{u-v coeff}
\end{equation}
we obtain

\begin{equation}
{\cal E}^{(2)}=\frac{M_{s}V}{\gamma }\sum\limits_{{\bf k}}\omega _{{\bf k}%
}c_{{\bf k}}^{\ast }c_{{\bf k}},\quad \omega _{{\bf k}}=\sqrt{{\cal A}_{{\bf %
k}}^{2}-{\cal B}_{{\bf k}}^{2}}.  \label{SWfreq}
\end{equation}
The spin-wave frequency, $\omega _{{\bf k}}$, in an explicit form is

\begin{eqnarray}
\omega _{{\bf k}} &=&\gamma \Big[H_{0}\cos (\theta _{H}-\theta
_{0})-H_{K}\sin ^{2}\theta _{0}  \nonumber \\
&&+4\pi M_{s}[1-G(k\tau )]\left( \frac{k_{y_{0}}}{k}\right) ^{2}+2\alpha
_{E}k^{2}\Big]^{1/2}  \nonumber \\
&&\times \left[ H_{0}\cos (\theta _{H}-\theta _{0})+4\pi M_{s}G(k\tau )%
\right] ^{1/2}.  \label{sw spectrum}
\end{eqnarray}

\subsection{Spin-wave interactions}

The interaction energy can be represented in the form:

\begin{eqnarray}
&&\frac{{\cal E}_{int}}{\left( M_{s}V/\gamma \right) }=%
\mathop{\displaystyle\sum}%
\limits_{{\bf 1},{\bf 2},{\bf 3}}\Big[\frac{\Psi _{1}({\bf 1},{\bf 2},{\bf 3}%
)}{3}c_{{\bf 1}}c_{{\bf 2}}c_{{\bf 3}}  \label{energy interactions} \\
&&+\Psi _{2}({\bf 1},{\bf 2};-{\bf 3})c_{{\bf 1}}c_{{\bf 2}}c_{-{\bf 3}%
}^{\ast }+{\rm c.c.}\Big]\Delta ({\bf 1}+{\bf 2}+{\bf 3})  \nonumber \\
&&+{\frac{1}{2}}\sum_{{\bf 1},{\bf 2},{\bf 3},{\bf 4}}\Phi ({{\bf 1},{\bf 2},%
{\bf 3},{\bf 4}})c_{{\bf 1}}^{\ast }c_{{\bf 2}}^{\ast }c_{{\bf 3}}c_{{\bf 4}%
}\Delta ({\bf 1}+{\bf 2}-{\bf 3}{-{\bf 4}}).  \nonumber
\end{eqnarray}
Here for simplicity we use the following notations: ${\bf k}_{1}\equiv {\bf 1%
}$, ${\bf k}_{2}\equiv {\bf 2}$, etc., for example, ${\bf k}_{1}+{\bf k}_{2}+%
{\bf k}_{3}\equiv {\bf 1}+{\bf 2}+{\bf 3}$. The three-wave interaction
amplitudes are

\begin{eqnarray}
\Psi _{1}({\bf 1},{\bf 2},{\bf 3}) &=&\frac{1}{2}\{\psi ({\bf 1})(u_{{\bf 1}%
}+v_{{\bf 1}})(u_{{\bf 2}}v_{{\bf 3}}+u_{{\bf 3}}v_{{\bf 2}})  \nonumber \\
&&+\psi ({\bf 2})(v_{{\bf 1}}u_{{\bf 3}}+v_{{\bf 3}}u_{{\bf 1}})(u_{{\bf 2}%
}+v_{{\bf 2}})  \nonumber \\
&&+\psi ({\bf 3})(v_{{\bf 1}}u_{{\bf 2}}+v_{{\bf 2}}u_{{\bf 1}})(u_{{\bf 3}%
}+v_{{\bf 3}})\}.  \label{ccc}
\end{eqnarray}
and

\begin{eqnarray}
\Psi _{2}({\bf 1},{\bf 2};{\bf 3}) &=&\frac{1}{2}\{\psi ({\bf 1})(u_{{\bf 1}%
}+v_{{\bf 1}})(u_{{\bf 2}}u_{{\bf 3}}+v_{{\bf 2}}v_{{\bf 3}})  \nonumber \\
&&+\psi ({\bf 2})(u_{{\bf 2}}+v_{{\bf 2}})(u_{{\bf 1}}u_{{\bf 3}}+v_{{\bf 1}%
}v_{{\bf 3}})  \nonumber \\
&&+\psi ({\bf 3})(u_{{\bf 3}}+v_{{\bf 3}})(v_{{\bf 1}}u_{{\bf 2}}+v_{{\bf 2}%
}u_{{\bf 1}})\}.  \label{ccc+}
\end{eqnarray}
Here 
\begin{equation}
\psi ({\bf k})=-\frac{\gamma }{\sqrt{2}}\left( \frac{H_{K}}{2}\sin 2\theta
_{0}+4\pi M_{s}[1-G(k\tau )]\frac{k_{y_{0}}k_{z_{0}}}{k^{2}}\right) .
\label{psi 3}
\end{equation}

The four-wave interaction amplitude can be expressed as:

\begin{equation}
\Phi ({{\bf 1},{\bf 2},{\bf 3},{\bf 4}})=\Phi _{0}({\bf 1},{\bf 2},{\bf 3},%
{\bf 4})+\Phi _{s}({\bf 1},{\bf 2},{\bf 3},{\bf 4})+\Phi _{Q}({\bf 1},{\bf 2}%
,{\bf 3},{\bf 4}),  \label{4 ampl}
\end{equation}
where

\begin{eqnarray}
\Phi _{0} &=&[\gamma H_{0}\cos (\theta _{H}-\theta _{0})+\gamma H_{K}\cos
^{2}\theta _{0}]  \nonumber \\
&&\times \Big\{u_{{\bf 1}}u_{{\bf 2}}v_{{\bf 3}}v_{{\bf 4}}+v_{{\bf 1}}v_{%
{\bf 2}}u_{{\bf 3}}u_{{\bf 4}}  \nonumber \\
&&-\frac{1}{2}[(u_{{\bf 1}}u_{{\bf 2}}+v_{{\bf 1}}v_{{\bf 2}})(u_{{\bf 3}}v_{%
{\bf 4}}+v_{{\bf 3}}u_{{\bf 4}})  \nonumber \\
&&+(u_{{\bf 1}}v_{{\bf 2}}+v_{{\bf 1}}u_{{\bf 2}})(u_{{\bf 3}}u_{{\bf 4}}+v_{%
{\bf 3}}v_{{\bf 4}})]\Big\},  \label{F0}
\end{eqnarray}

\begin{eqnarray}
\Phi _{s} &=&\frac{1}{4}\left[ {\cal P}({\bf 1})+{\cal P}({\bf 2})+{\cal P}(%
{\bf 3})+{\cal P}({\bf 4})\right]  \nonumber \\
&&\times \lbrack (u_{{\bf 1}}u_{{\bf 2}}-v_{{\bf 1}}v_{{\bf 2}})(v_{{\bf 3}%
}v_{{\bf 4}}-u_{{\bf 3}}u_{{\bf 4}})  \nonumber \\
&&-\left( u_{{\bf 1}}v_{{\bf 2}}+v_{{\bf 1}}u_{{\bf 2}}\right) (v_{{\bf 3}%
}v_{{\bf 4}}+u_{{\bf 3}}u_{{\bf 4}})  \nonumber \\
&&-(u_{{\bf 1}}u_{{\bf 2}}+v_{{\bf 1}}v_{{\bf 2}})(v_{{\bf 3}}u_{{\bf 4}}+u_{%
{\bf 3}}v_{{\bf 4}})  \nonumber \\
&&-\frac{1}{2}(v_{{\bf 1}}u_{{\bf 2}}+u_{{\bf 1}}v_{{\bf 2}})(v_{{\bf 3}}u_{%
{\bf 4}}-u_{{\bf 3}}v_{{\bf 4}})],  \label{simm ampl}
\end{eqnarray}

\begin{eqnarray}
\Phi _{Q} &=&[{\cal Q}({\bf 1+2})+{\cal Q}({\bf 3+4})](u_{{\bf 1}}u_{{\bf 2}%
}u_{{\bf 3}}u_{{\bf 4}}+v_{{\bf 1}}v_{{\bf 2}}v_{{\bf 3}}v_{{\bf 4}}) 
\nonumber \\
&&+[{\cal Q}({\bf 1+4})+{\cal Q}({\bf 2+3})](u_{{\bf 1}}v_{{\bf 2}}u_{{\bf 3}%
}v_{{\bf 4}}+v_{{\bf 1}}u_{{\bf 2}}v_{{\bf 3}}u_{{\bf 4}})  \nonumber \\
&&+[{\cal Q}({\bf 1+3})+{\cal Q}({\bf 2+4})](u_{{\bf 1}}v_{{\bf 2}}v_{{\bf 3}%
}u_{{\bf 4}}+v_{{\bf 1}}u_{{\bf 2}}u_{{\bf 3}}v_{{\bf 4}}),  \label{def F Q}
\end{eqnarray}

\begin{equation}
\frac{{\cal P}({\bf k})}{\gamma }\equiv \alpha _{E}k^{2}-\frac{H_{K}}{2}\sin
^{2}\theta _{0}+2\pi M_{s}[1-G(k\tau )]\left( \frac{k_{y_{0}}}{k}\right)
^{2},  \label{def P}
\end{equation}
and 
\begin{equation}
\frac{{\cal Q}({\bf k})}{\gamma }\equiv \alpha _{E}k^{2}-\frac{H_{K}}{2}\cos
^{2}\theta _{0}+2\pi M_{s}[1-G(k\tau )]\left( \frac{k_{z_{0}}}{k}\right)
^{2}.  \label{def R}
\end{equation}

\subsection{Effective four-wave interactions}

Unitary transformation (see, \cite{ShiSaf}) makes it possible to eliminate
forbidden three-magnon interaction terms in Eq.(\ref{energy interactions})
and obtain effective interaction terms. As a result we have the following
spin-wave energy

\begin{eqnarray}
&&{\cal E}/\left( \frac{M_{s}V}{\gamma }\right) =\sum\limits_{{\bf k}}\omega
_{{\bf k}}c_{{\bf k}}^{\ast }c_{{\bf k}}  \label{energy diag} \\
&&+{\frac{1}{2}}\sum_{{\bf 1},{\bf 2},{\bf 3},{\bf 4}}\widetilde{\Phi }({%
{\bf 1},{\bf 2},{\bf 3},{\bf 4}})c_{{\bf 1}}^{\ast }c_{{\bf 2}}^{\ast }c_{%
{\bf 3}}c_{{\bf 4}}\Delta ({\bf 1}+{\bf 2}-{\bf 3}{-{\bf 4}}),  \nonumber
\end{eqnarray}
where

\begin{equation}
\widetilde{\Phi }({{\bf 1},{\bf 2},{\bf 3},{\bf 4}})={\Phi }({{\bf 1},{\bf 2}%
,{\bf 3},{\bf 4}})+\Phi _{1}({{\bf 1},{\bf 2},{\bf 3},{\bf 4}})+\Phi _{2}({%
{\bf 1},{\bf 2},{\bf 3},{\bf 4}}),  \label{4-magnon amplitude}
\end{equation}
\begin{eqnarray}
&&\Phi _{1}({{\bf 1},{\bf 2},{\bf 3},{\bf 4}})=-\Psi _{1}({\bf 1}{,{\bf 2},-}%
{\bf 1}-{\bf 2})\Psi _{1}({\bf 3}{,{\bf 4},-{\bf 3}-{\bf 4}})  \nonumber \\
&&\quad \times \left( \frac{1}{\omega _{{\bf 1}}+\omega _{{\bf 2}}+\omega _{-%
{\bf 1}-{\bf 2}}}+\frac{1}{\omega _{{\bf 3}}+\omega _{{\bf 4}}+\omega _{-%
{\bf 3}-{\bf 4}}}\right) ,  \label{F1 final}
\end{eqnarray}
and 
\begin{eqnarray}
&&\Phi _{2}({\bf 1}{,{\bf 2},}{\bf 3}{,{\bf 4}})=\Psi _{2}({\bf 1}{,{\bf 2},%
{\bf 1}+{\bf 2}})\Psi _{2}({\bf 3}{,{\bf 4},{\bf 3}+{\bf 4}})  \nonumber \\
&&\qquad \times \left( \frac{1}{\omega _{{\bf 1}}+\omega _{{\bf 2}}-\omega _{%
{\bf 1}+{\bf 2}}}+\frac{1}{\omega _{{\bf 3}}+\omega _{{\bf 4}}-\omega _{{\bf %
3}+{\bf 4}}}\right)  \nonumber \\
&&\quad -\Psi _{2}({\bf 1}{,{\bf 3}-{\bf 1},{\bf 3}})\Psi _{2}({\bf 4}{,{\bf %
2}-{\bf 4},{\bf 2}})  \nonumber \\
&&\qquad \times \left( \frac{1}{\omega _{{\bf 1}}+\omega _{{\bf 3-1}}-\omega
_{{\bf 3}}}+\frac{1}{\omega _{{\bf 4}}+\omega _{{\bf 2}-{\bf 4}}-\omega _{%
{\bf 2}}}\right)  \nonumber \\
&&\quad -\Psi _{2}({\bf 1}{,{\bf 4}-{\bf 1},{\bf 4}})\Psi _{2}({\bf 3}{,{\bf %
2}-{\bf 3},{\bf 2}})  \nonumber \\
&&\qquad \times \left( \frac{1}{\omega _{{\bf 1}}+\omega _{{\bf 4-1}}-\omega
_{{\bf 4}}}+\frac{1}{\omega _{{\bf 3}}+\omega _{{\bf 2-3}}-\omega _{{\bf 2}}}%
\right)  \nonumber \\
&&\quad -\Psi _{2}({\bf 2}{,{\bf 3}-{\bf 2},{\bf 3}})\Psi _{2}({\bf 4}{,{\bf %
1}-{\bf 4},{\bf 1}})  \nonumber \\
&&\qquad \times \left( \frac{1}{\omega _{{\bf 2}}+\omega _{{\bf 3}-{\bf 2}%
}-\omega _{{\bf 3}}}+\frac{1}{\omega _{{\bf 4}}+\omega _{{\bf 1}-{\bf 4}%
}-\omega _{{\bf 1}}}\right)  \nonumber \\
&&\quad -\Psi _{2}({\bf 2}{,{\bf 4}-{\bf 2},{\bf 4}})\Psi _{2}({\bf 3}{,{\bf %
1}-{\bf 3},{\bf 1}})  \nonumber \\
&&\qquad \times \left( \frac{1}{\omega _{{\bf 2}}+\omega _{{\bf 4}-{\bf 2}%
}-\omega _{{\bf 4}}}+\frac{1}{\omega _{{\bf 3}}+\omega _{{\bf 1-3}}-\omega _{%
{\bf 1}}}\right) .  \label{F2 final}
\end{eqnarray}

The energy (\ref{energy diag}) together with the Hamilton's equations of
motion for complex spin-wave variables

\begin{equation}
\left( \frac{d}{dt}+\eta _{{\bf k}}\right) c_{{\bf k}}=-i\left( \frac{\gamma 
}{M_{s}V}\right) \frac{\partial {\cal E}}{\partial c_{{\bf k}}^{\ast }},
\label{Ham equ}
\end{equation}%
supplemented by the (microscopic) relaxation rate $\eta _{{\bf k}}$,
represent the basis for magnetization dynamics modeling in the ${\bf k}$%
-space. Calculating $c_{{\bf k}}(t)$, $c_{{\bf k}}^{\ast }(t)$, we can find
magnetization deviations $a_{{\bf k}}(t)$, $a_{{\bf k}}^{\ast }(t)$ with the
help of back transformation (\ref{u-v transform}) and finally with Eqs. (\ref%
{mx1})-(\ref{mz1}), the averaged 
\begin{equation}
\left\langle {\bf m}({\bf r},t)\right\rangle =\frac{1}{L_{y}L_{z}}%
\displaystyle\int %
\limits_{0}^{L_{y}}dy%
\displaystyle\int %
\limits_{0}^{L_{z}}dz\ {\bf m}({\bf r},t)={\bf m}_{0}(t),  \label{averaged m}
\end{equation}%
which gives a measure of non-uniform magnetization motions in the system. In
general, $\left\vert {\bf m}_{0}\right\vert <1$ and only in the case of
coherent spin motion $\left\vert {\bf m}_{0}\right\vert =1$.

\subsection{Example}

For simplicity we shall consider the uniform magnetization precession
interacting with one spin-wave pair with ${\bf k}=(0,0,k)$ along $\widehat{%
{\bf z}}_{0}$. In this case the energy (Hamiltonian) of the system can be
reduced to the form:

\begin{equation}
{\cal H}={\cal H}_{0}+{\cal H}_{int}+{\cal H}_{p},  \label{Ham pair}
\end{equation}
where

\begin{equation}
{\cal H}_{0}/\left( \frac{M_{s}V}{\gamma }\right) =\omega _{0}c_{0}^{\ast
}c_{0}+\omega _{{\bf k}}(c_{{\bf k}}^{\ast }c_{{\bf k}}+c_{-{\bf k}}^{\ast
}c_{-{\bf k}})  \label{Ham quadrat}
\end{equation}
describes the uniform precession and spin-wave pair,

\begin{eqnarray}
&&\frac{{\cal H}_{int}}{(M_{s}V/\gamma )}=\frac{\Phi _{00}}{2}c_{0}^{\ast
}c_{0}^{\ast }c_{0}c_{0}+2\Phi _{0{\bf k}}c_{0}^{\ast }c_{0}(c_{{\bf k}%
}^{\ast }c_{{\bf k}}+c_{-{\bf k}}^{\ast }c_{-{\bf k}})  \label{inter ham} \\
&&+\frac{\Phi _{{\bf kk}}}{2}(c_{{\bf k}}^{\ast }c_{{\bf k}}c_{{\bf k}%
}^{\ast }c_{{\bf k}}+c_{-{\bf k}}^{\ast }c_{-{\bf k}}c_{-{\bf k}}^{\ast }c_{-%
{\bf k}})+2\Phi _{{\bf k,-k}}c_{{\bf k}}^{\ast }c_{{\bf k}}c_{-{\bf k}%
}^{\ast }c_{-{\bf k}}  \nonumber
\end{eqnarray}
describes nonlinear interactions in the system, $\Phi _{00}\equiv \widetilde{%
\Phi }(0,0,0,0)$, $\Phi _{0{\bf k}}\equiv \widetilde{\Phi }({\bf k},0,{\bf k}%
,0)$, $\Phi _{{\bf kk}}\equiv \widetilde{\Phi }({\bf k},{\bf k},{\bf k},{\bf %
k})$, $\Phi _{{\bf k,-k}}\equiv \widetilde{\Phi }({\bf k},-{\bf k},{\bf k},-%
{\bf k})$, and

\begin{equation}
{\cal H}_{p}/\left( \frac{M_{s}V}{\gamma }\right) =\Phi _{p}\left(
c_{0}c_{0}c_{{\bf k}}^{\ast }c_{-{\bf k}}^{\ast }+c_{0}^{\ast }c_{0}^{\ast
}c_{{\bf k}}c_{-{\bf k}}\right)  \label{pumping}
\end{equation}
describes the spin-wave pair excitation by the uniform precession, $\Phi
_{p}\equiv \widetilde{\Phi }({\bf k},-{\bf k},0,0)$.

The uniform precession and spin-wave pair dynamics are defined by

\begin{mathletters}
\begin{eqnarray}
\left( \frac{d}{dt}{+}\eta _{0}\right) c_{0} &=&-i\widetilde{\omega }%
_{0}c_{0}-2i\Phi _{p}c_{{\bf k}}c_{-{\bf k}}c_{0}^{\ast },  \label{c0 dyn} \\
\left( \frac{d}{dt}{+}\eta _{{\bf k}}\right) c_{{\bf k}} &=&-i\widetilde{%
\omega }_{{\bf k}}c_{{\bf k}}-i\Phi _{p}(c_{0})^{2}c_{-{\bf k}}^{\ast },
\label{ck dyn}
\end{eqnarray}
where

\end{mathletters}
\begin{eqnarray}
\widetilde{\omega }_{0} &=&\omega _{0}+\widetilde{\Phi }_{00}|c_{0}|^{2}+2%
\Phi _{0{\bf k}}(|c_{{\bf k}}|^{2}+|c_{-{\bf k}}|^{2}),  \label{wo non} \\
\widetilde{\omega }_{{\bf k}} &=&\omega _{{\bf k}}+2\widetilde{\Phi }_{0{\bf %
k}}|c_{0}|^{2}+\Phi _{{\bf kk}}|c_{{\bf k}}|^{2}+2\Phi _{{\bf k,-k}}|c_{-%
{\bf k}}|^{2}  \label{wk non}
\end{eqnarray}
are nonlinear frequencies and ${\eta _{0}}$, $\eta _{{\bf k}}$ are the
relaxation rates.

From the energy symmetry we have $c_{{\bf k}}=c_{-{\bf k}}$ (see, also \cite%
{Safpair}). Thus, Eqs.(\ref{c0 dyn})-(\ref{wk non}) represent a
self-consistent nonlinear theory of magnetization reversal with just two
independent complex variables. Simple analysis for both $c_{{\bf k}}(t)$ and 
$c_{-{\bf k}}^{\ast }(t)\propto \exp (\kappa t)$, where $\kappa $ is an
increment of instability, gives:

\begin{equation}
\kappa ={-}\eta _{{\bf k}}+\sqrt{\left\vert \Phi _{p}c_{0}^{2}(0)\right\vert
^{2}+\left( \widetilde{\omega }_{{\bf k}}-\widetilde{\omega }_{0}\right) ^{2}%
}.  \label{increment}
\end{equation}%
This formula is similar to that obtained in Ref.\cite{cpatton} for
parametric instabilities. The difference is that the resonance condition
includes nonlinear frequencies: $2\widetilde{\omega }_{0}=\widetilde{\omega }%
_{{\bf k}}+\widetilde{\omega }_{-{\bf k}}$ (all spin waves out of this
equality can be included into a thermal bath) and $\widetilde{\omega }_{0}$
is the uniform precession frequency (not a driving field frequency). The
onset of instability is defined by $\left\vert \Phi
_{p}c_{0}^{2}(0)\right\vert \geq \eta _{{\bf k}}$. Taking $m_{x_{0}}(0)=0$,
from Eqs.(\ref{my}), (\ref{mz}) and (\ref{u-v transform}) we can find $%
c_{0}(0)$ and rewrite this criterion as

\begin{equation}
\theta \geq (u_{0}+v_{0})\sqrt{2\eta _{{\bf k}}/|\Phi _{p}|},
\label{criterion theta}
\end{equation}%
where $\theta \equiv \tan ^{-1}(m_{y_{0}}/m_{z_{0}})$ is the initial
deviation angle from the equilibrium direction $z_{0}$. Here the parameters $%
u_{0}$, $v_{0}$ and $\Phi _{p}$ can be directly calculated using Eqs. (\ref%
{u-v transform}), (\ref{pumping}). The damping $\eta _{{\bf k}}$\ can be
estimated microscopically \cite{nonrelax}, \cite{safbertrelax}.

Fig.2 shows two different evolutions in magnetic system calculated by Eqs.(%
\ref{c0 dyn}), (\ref{ck dyn}). Spin-wave pair excitation in Fig.2a is not
sufficiently strong during the switching process to affect the uniform
magnetization dynamics. The averaged $\left\langle {\bf m}({\bf r}%
)\right\rangle ={\bf m}_{0}$, which gives a measure of non-uniform
magnetization motions in the system is relatively small ($1-\left\vert {\bf m%
}_{0}\right\vert \lesssim 0.05$). Fig.2b demonstrates a strong excitation of
spin-wave instability by uniform precession and substantial increase of the
magnetization reversal rate. In this case $\left\vert {\bf m}_{0}\right\vert 
$ reaches $\simeq 0.6$. For stronger coupling ($|\Phi _{p}|/\eta _{{\bf k}%
}>21$) we observed beating between the uniform precession and the spin-wave
pair. In this case it is necessary to include into consideration the
excitation of another resonance spin-wave pairs.

\section{Discussion}

We have considered two principally different scenarios of magnetization
reversal. In the first one the coherent rotation of all spins is the
principal motion in the system. The magnetization dynamics is defined by Eq.
(\ref{basic equation 1}), which is reduced to Bloch-Bloembergen form in the
case of small magnetization motions.

In the second scenario we have considered an ultra-thing ferromagnetic film
with large dimensions in the plane. In this case the role of plane
boundaries is negligible and the most convenient technique to describe the
non-uniform spin motions is their spin-wave representation in the ${\bf k}$%
-space. Taking into account linear spin-wave modes and their scattering, we
have constructed a nonlinear self-consistent theory of magnetization
reversal as a decay of uniform magnetization precession and nonlinear
excitation of spin-wave pairs. This theory includes an effective energy (\ref%
{energy diag}) and dynamic equations (\ref{Ham equ}). The most important
spin-wave modes are defined by the resonance condition $2\widetilde{\omega }%
_{0}=\widetilde{\omega }_{{\bf k}}+\widetilde{\omega }_{-{\bf k}}$ (similar
to Suhl's second order instability). The excitation of all spin waves out of
this resonance is small and therefore they can be considered as a part of a
thermal bath. In the simple example we have demonstrated that strongly
excited spin-wave instability can increase substantially the magnetization
switching rate.

Note that both scenarios are consistent which each other: in the case of $%
\left\vert {\bf m}_{0}\right\vert \rightarrow 1$ Eqs.(\ref{Ham equ}) can be
reduced to the case of a nonlinear oscillator equation considered in
scenario 1. We also emphasize that the relaxation rates of uniform
precession and spin waves in both scenarios can be estimated from
microscopic physics.

\section*{Acknowledgments}

The author wish to thank H. N. Bertram for many stimulating and critical
discussions. I also greatly appreciate helpful discussions with H. Suhl, C.
E. Patton, T. J. Silva, P. Kabos, and J. P. Nibarger. This work was
supported in part by the National Institute of Standards and Technology's
Nanotechnology Initiative and partly supported by matching funds from the
Center for Magnetic Recording Research at the University of California - San
Diego and CMRR incorporated sponsor accounts.

\newpage Figure Captions

Fig.1 Equilibrium coordinates in the film plane.

Fig.2 Time evolution of relative absolute amplitudes: 1- the uniform
precession $|c_{0}(t)|$ without spin-wave excitation, 2 - the uniform
precession $|c_{0}(t)|$\ with spin-wave excitation, 3 - excited spin waves $%
|c_{k}(t)|$. Curve \# 4 describes $\left\vert {\bf m}_{0}(t)\right\vert $.
a) $|\Phi _{p}|/\eta _{{\bf k}}=9$, b) $|\Phi _{p}|/\eta _{{\bf k}}=21$. The
experimental conditions correspond to Fig.2 in Ref. \cite{SilvaKabosPufall}.

\end{document}